\documentclass[showkeys]{revtex4}

\usepackage{epsfig}

\usepackage{amssymb}

\usepackage{lineno,hyperref}
\usepackage{cancel}
\usepackage{ulem}

\usepackage{graphicx}
\usepackage{dcolumn}
\usepackage{bm}
\usepackage{amsmath}

\begin{document}

\title{Systematic Oil Flow Modeling in the Quasi-3D Approximation Yields Additional Terms that Allows for Variable Cross-Section Area Tubing}
\author{Edval J.\ P.\ Santos}
\affiliation{Laboratory for Devices and
Nanostructures -- Nanoscale Engineering Group,
Universidade Federal de Pernambuco, Recife-PE, Brazil\\
E-mail: edval@ee.ufpe.br or e.santos@expressmail.dk.
}

\begin{abstract}
A systematic model development for oil flow in quasi-3D (1D + 2D) is presented.  Our approach provides a unified modeling scheme. Besides, additional terms are obtained, which allows for tubing area variation along the flow direction.  The area variation can be modeled as analytic function or random fluctuation, as it could be the result of deposits or tubing internal surface roughness.  The proposed approach can be used to obtain analytic solutions which provide physical insight into the phenomena under scrutiny, including the validation of software tools, sensor development and sensor placement.  One starts from conservation laws as given by kinetic theory and applies the transverse averaging technique (TAT) to extract the one-dimensional approximation in formal grounds.  To demonstrate its application, the steady-state Ramey's model,  the Hasan's transient model and a simple two-phase model are generated from the obtained equations.

\end{abstract}



\maketitle

\section{Introduction}
\label{sec:Introduction}

Many models have been developed to describe oil and gas flow in production wells or injection fluid in artificial lift wells in terms of flowrate, pressure profile, and temperature profile.  A pioneer work in such development is the one dimensional phenomenological model  for steady-state flow in injection wells reported by Henry J. Ramey Jr.~\cite{Ramey1962} in 1962.  Later this model was improved for transient flow.  With the advancement of computers, 3D models have been developed and implemented as computer software.  Despite the sophistication of computer tools, analytic models stills plays a role to offer insight into physical problems, to help in tool validation, sensor development and sensor placement~\cite{HKW1997,CC2005,Izgec2008,HK2012,SUHPG2013,FINALREPORT,SL2018,SYCLZMSZF2017,SYLLLC2018,SP2016,SS2010,SS2013,SS2014,SS2019}.

In the oil and gas industry, balance of mass, momentum or energy are applied to a fluid parcel volume to build models for the extracted fluid flow in production wells or the injection fluid in artificial lift wells.  The model equations are combined with equations of state and empirical equations.  Depending on the model, many other ancillary parameters need to be measured or modeled~\cite{JM2004}.  Full 3D coupled models are used during well drilling to monitor pressure surge or swab to keep the drilling process within the safe window to avoid well fracture or blowout.  During production, the model is used for keeping the well in production, deciding about treatment, or deciding about shut-in for maintenance, safety precaution  or economical reasons~\cite{Ramey1962,HKW1997,CC2005,Izgec2008,HK2012,SUHPG2013,FINALREPORT,SL2018,SYCLZMSZF2017,SYLLLC2018}.

Extracted fluid is a combination of many components, such as: oil, water, gas, and sediments.  Depending on the viscosity of the oil, different viscosity models apply, such as: power law fluid, Bingham fluid, yield power law fluid, Casson fluid, among others.  Range of crude oil density~\cite{API} is shown in Table~\ref{tab:oildensity}.

\begin{table}[h]
\begin{center}
\caption{Crude oil density.}
\begin{tabular}{l|c|c}
\hline
            & API gravity & Kg/m$^3$         \\
\hline      
Light       & $>$ 31.1      & $<$ 870        \\
Medium      & 22.3 - 31.1   & 870 - 920      \\
Heavy       & 10 - 22.3     & 920 - 1000     \\
Extra heavy & $<$ 10        & $>$ 1000       \\
\hline
\end{tabular}
\label{tab:oildensity}
\end{center}
\end{table}

The simplest model of the extracted fluid is a mixture of ideal gas and  incompressible Newtonian liquid.  A two-phase flow.  Considering the two-phase mixture, the simplest approximation for the flow density, $\rho_F$, is a linear combination.    Thus, the density of the fluid can be estimated as follows.  

\begin{equation}
\rho_F= \rho_LH_L + \rho_G(1-H_L)
\end{equation}
in which, $H_L$ is the liquid fraction also known as liquid holdup.

An equivalent expression can be written for the fluid velocity.

\begin{equation}
\rho_F \vec{u}= \rho_L\vec{u}_LH_L + \rho_G\vec{u}_G(1-H_L)
\end{equation}

In the simplest model, liquid and gas velocities are equal.  This is the basis of the  Homogeneous Equilibrium Model (HEM).  Unequal liquid and gas velocities results in phase sliding.  As oil and gas move up to the wellhead as a function of pressure profile, it exchanges thermal energy with the surroundings, resulting in a variable temperature profile in the wellbore.  Variable temperature affects oil and gas density, and many other properties of the fluid.  In return, this causes a variation of the pressure profile, and the flow regime.  One says, the flow equations describing such phenomena are coupled.  Besides, along the tubing, the flow regime can vary, ranging from plug flow to laminar flow to turbulent flow. Plug flow is a single phase smooth flow.  A characteristic of two-phase flow is the presence of one or more interfaces separating the phases, a few examples of flow regime with approximate cylindrical symmetry are~\cite{IH2006,DP1999}:

\begin{itemize}
\item Stratified --  at low gas and liquid flowrates, phase separated, multicomponent smooth flow.
\item Stratified/wavy --  at higher gas and liquid flowrates, alternates between liquid and gas.
\item Slug -- at medium gas and liquid flowrates, large bubbles.
\item Bubbly --  at high liquid flowrates, bubbles of gas in the liquid phase.
\item Annular -- at high gas flowrates, gas with liquid wetting the pipe wall.
\end{itemize}

To solve the full 3D coupled problem of a multiphase flow is a difficult task even for multiphysics tools.  Approximate methods based on averaging techniques are useful to offer insight and to validate software tools.  There are many approximate methods based on area averaging, such as: Homogeneous Equilibrium Model (HEM), Separated Flow Model (SFM), Drift Flux Model (DFM), and Two Fluid Model (TFM).  In the Homogeneous Equilibrium Model is assumed that fluid phases are in equilibrium with each other, so that any differences in velocity, pressure or temperature rapidly disappear resulting in a single phase pseudo fluid.  In the Separated Flow Model, the phases do not necessarily have the same velocity.  Thus, there is a relative or drift velocity of the lighter phase over the heavier one, i.e, phase slip can occur with equal pressure and temperature.  The Drift Flux Model is a particular case of the separated flow model in which the drift velocity is written as a function of external forces and fluid properties.  The Two Fluid Model is the most complex model, each phase can achieve its own velocity, pressure and temperature profile, and is treated separately~\cite{IH2006,DP1999}.

In kinetic theory, conservation laws play a key role in describing system behavior.  Frequently, conservation of particles, conservation of momentum and conservation of energy equations are used to provide insight into many physical problems.  Phenomena ranging from flow of charge carriers in a semiconductor device to flow of oil and gas in a well can be described within such theoretical framework~\cite{IH2006,DP1999,RLL1990,BS2007,BS2011}.

In a previous work, we have demonstrated how to break the problem into 1D + 2D dimensions by applying the transverse averaging technique (TAT), and we have used this approximation to study the nonuniformly doped or shaped $p$-$n$ junction diodes.  In the occasion the TAT technique was applied to the conservation of particles equation for holes and electrons~\cite{BS2007,BS2011}.  In TAT, the cross-section area is not necessarily held constant.   In this paper the TAT technique is extended to tensor quantities and applied to the conservation of mass, conservation of momentum and conservation of energy equations from kinetic theory.   Thus, the one-dimensional approximation of the three equations is set on formal grounds without assuming constant cross-section area.   As an application example, a set of equations is obtained to describe flow in an oil well.  The area variation can be modeled as analytic function or random fluctuation, as it could be the result of deposits or tubing internal surface roughness. 

The goal of this approach is not to replace sophisticated models, but to provide a systematic model construction approach in 1D + 2D which can be used to provide insight into oil flow or to test software tools.  In our approach, the conservation equations are simplified with the transverse averaging technique (TAT) to provide equations for oil flow.  Under appropriate assumptions, the resulting equations are equivalent to equations obtained by studying a control parcel volume~\cite{Ramey1962,HKW1997}.  The paper is divided into six sections.  This introduction is the first.  Next, the conservation equations in fluids are discussed.  In the third section, the TAT approximation is applied.  In the fourth section, the set of equations for oil flow is obtained.  In the fifth section, application examples are presented.  Finally, the conclusions.


\section{Conservation equations in fluids}
\label{sec:conservequations}

To build a mathematical model for the flow, the fluid is divided into small parts or parcels with identical mass, $m$, which are small enough to justify the continuum approximation and large enough as compared to the molecules.  Thus, the exact number of parcels, $N$, is not well defined~\cite{IH2006,DP1999,SC2008,Chaudhry2014,RK2016}.

The distribution of parcels or particles in the phase space is given by the probability density function, also named distribution function or particle density function, $F_B(\vec{r},\vec{v},t)= Nf_B(\vec{r},\vec{v},t)$.  It is the probability density of finding a parcel at position between $\vec{r}$ and $\vec{r} + d\vec{r}$, with velocity between $\vec{v}$ and  $\vec{v} + d\vec{v}$, at instant $t$.  From the probability density function one can extract properties of the system, such as fluid density, $\rho_F$, momentum density, $\rho_F\vec{u}$, total pressure, $\overline{\overline{P}}$, total flux of kinetic energy density, $\Phi_q$, and total kinetic energy density, $e_K$.  This is also named Boltzmann statistical average~\cite{IH2006,DP1999,RLL1990}.

\begin{eqnarray}
\rho_F(\vec{r},t)&=& \int m\,F_B(\vec{r},\vec{v},t)\,d\vec{v}\label{eq:density}\\
\rho_F(\vec{r},t)\,\vec{u}(\vec{r},t)&=& \int m\vec{v}F_B(\vec{r},\vec{v},t)\,d\vec{v}\\
\overline{\overline{P}}(\vec{r},t)&=& \int m\,\overline{\overline{vv}}\,F_B(\vec{r},\vec{v},t)\,d\vec{v}\label{eq:pressure}\\
e_K(\vec{r},t)&=&  \int {1\over 2}m\left(\vec{v}\cdot\vec{v}\right)F_B(\vec{r},\vec{v},t)\,d\vec{v}\label{eq:energy}\\
\Phi_q(\vec{r},t)&=& \int {1\over 2}m\left(\vec{v}\cdot\vec{v}\right)\vec{v}F_B(\vec{r},\vec{v},t)\,d\vec{v}\label{eq:heat} 
\end{eqnarray}
in which, $\vec{u}$ is the parcel average velocity, $\overline{\overline{P}}$ is the pressure tensor, and $\overline{\overline{vv}}= \vec{v}\,\vec{v}^T$ is the tensor dyadic product.

The parcel velocity, $\vec{v}$, can be written in terms of the average velocity, $\vec{u}$, which is the fluid velocity.

\begin{equation}
\vec{v}= \vec{u} + \vec{c}
\label{eq:averagevelocity}
\end{equation}
in which, $\vec{c}$, is the velocity of the parcel in a reference frame moving with the fluid at the average velocity.

In general, the dynamics of a set of classical particles is described by Boltzmann equation, from which one derives conservation equations for particles, momentum, and energy~\cite{RLL1990}.

\begin{itemize}

\item Conservation of mass density.

\begin{equation}
{\partial \rho_F\over \partial t} + \nabla\cdot \left(\rho_F\vec{u}\right)=  {\cal G}_N
\label{eq:conservmass}
\end{equation}
in which, ${\cal G}_N=  {\cal G} - {\cal R}$ is the net generation, ${\cal G}$ is related to the source or creation of parcels, and ${\cal R}$ is related to the sink or destruction of parcels.  

\item Conservation of momentum density.

\begin{equation}
{\partial (\rho_F\vec{u})\over \partial t} + \nabla\cdot \overline{\overline{P}}= \vec{f}_{ext}
\label{eq:conservmomentum}
\end{equation}
in which, $\overline{\overline{P}}$ is the total pressure tensor or total flux of momentum density.

\item Conservation of energy density.

\begin{equation}
{\partial e_K \over \partial t} + \nabla\cdot \Phi_q= \vec{f}_{ext}\cdot \vec{u}
\label{eq:conservenergy}
\end{equation}
in which, $e_K$ is the total kinetic energy density, and $\Phi_q$ is the total flux of kinetic energy density.

\end{itemize}

Rewriting the velocity in terms of the average, as shown in Equation~\ref{eq:averagevelocity}, and inserting in Equations~\ref{eq:pressure}, \ref{eq:energy}, and \ref{eq:heat}, one gets the following expressions.

\begin{eqnarray}
\overline{\overline{P}}&=&  \overline{\overline{p}} + \rho_F \overline{\overline{uu}}\label{eq:newpressure}\\
e_K&=& {U\over V} + {1\over 2}\rho_F \left(\vec{u}\cdot\vec{u}\right)\label{eq:newkinetic}\\
\Phi_q&=& \Phi_Q + e_K\vec{u} +\overline{\overline{p}}\cdot\vec{u}\label{eq:newheat}
\end{eqnarray}
in which, ${U\over V}=  {3\over 2}nk_BT$.

\subsection{Conservation of particles}

Defining the material or convective derivative, ${D\over Dt} \equiv {\partial \over \partial t} +  \vec{u}\cdot\nabla$, also known as the Eulerian operator.  The material derivative is the time variation of the quantity of interest as the fluid moves along a trajectory.  Considering the equation of conservation of particles, and using the identity $\nabla\cdot (\rho_F\vec{u})= \rho_F\nabla\cdot \vec{u} + \vec{u}\cdot(\nabla \rho_F) $, one can rewrite as follows~\cite{SC2008}.

\begin{equation}
{\partial \rho_F\over \partial t} + \nabla\cdot(\rho_F \vec{u})=  {D\rho_F\over Dt} + \rho_F\nabla\cdot \vec{u}= {\cal G}_N
\end{equation}
in which, ${\cal G}_N$ is the net generation term.  It can be used to model the reservoir.  Above the reservoir,  ${\cal G}_N= 0$.

The material derivative is related to the creation or destruction of mass.  For an incompressible fluid, $\nabla\cdot\vec{u}= 0$.  Above the reservoir the flow can be assumed free of sink or sources, ${\cal R}= 0$, ${\cal G}= 0$, the material derivative is zero.

\begin{equation}
{D\rho_F\over Dt}= 0
\end{equation}

\subsection{Conservation of momentum}

Replacing Equation~\ref{eq:newpressure} in the equation for the conservation of momentum, Equation~\ref{eq:conservmomentum}.

\begin{equation}
{\partial (\rho_F\,\vec{u})\over \partial t} + \nabla\cdot   \rho_F \overline{\overline{uu}}= - \nabla\cdot \overline{\overline{p}} + \vec{f}_{ext}
\end{equation}
in which, $\overline{\overline{uu}}= \vec{u}\,\vec{u}^T$ is the tensor dyadic product, and $ \left.\nabla\cdot \overline{\overline{p}}\right|_i= \sum_j{\partial p_{ij}\over \partial x_j}$.

Replacing the identity $\nabla\cdot (\rho_F\vec{u}\vec{u}^T)= \vec{u}(\nabla\cdot \rho_F\vec{u}) + (\rho_F\vec{u}\cdot \nabla)\vec{u}$, one gets a new equation for the conservation of momentum.

\begin{equation}
{\partial (\rho_F\vec{u})\over \partial t} + \vec{u}(\nabla\cdot \rho_F\vec{u}) + (\rho_F\vec{u}\cdot \nabla)\vec{u}=  -\nabla\cdot \overline{\overline{p}} + \vec{f}_{ext}
\label{eq:momconserv04}
\end{equation}

The pressure tensor can be replaced by the Cauchy stress tensor, $\overline{\overline{\sigma}}$, as follows~\cite{RK2016}.

\begin{equation}
-\overline{\overline{p}}= \overline{\overline{\sigma}}=
\left[
\begin{array}{ccc}
\sigma_{11} &  \tau_{12}   & \tau_{13} \\
\tau_{21}   &  \sigma_{22} & \tau_{23} \\
\tau_{31}   &  \tau_{32}   & \sigma_{33} \\
\end{array}
\right]
\end{equation}
in which, $\sigma_{ii}$ are the normal stresses and $\tau_{ij}$ are the shear or viscous stresses. 

If the external force per unit volume acting on the fluid is gravity, then $\vec{f}_{ext}= \rho_F \vec{g}$.  Combining the conservation of momentum equation with the conservation of mass.

\begin{equation}
\rho_F {\partial \vec{u}\over \partial t} + (\rho_F\vec{u}\cdot\nabla) \vec{u} + \vec{u}({\cal G} - {\cal R})= \nabla\cdot \overline{\overline{\sigma}} + \rho_F \vec{g}
\end{equation}

For a fluid free of sink or sources, ${\cal R}= 0$, ${\cal G}= 0$, one gets the Cauchy momentum equation.  

\begin{equation}
\rho_F {\partial \vec{u}\over \partial t} + (\rho_F\vec{u}\cdot\nabla) \vec{u}= \nabla\cdot \overline{\overline{\sigma}} + \rho_F \vec{g}
\label{eq:cauchy}
\end{equation}

For a Newtonian fluid, the stress tensor is a combination of hydrostatic pressure and shear stress proportional to the shear rate.

\begin{equation}
\sigma_{ij}= -p\delta_{ij} + \beta(\nabla\cdot\vec{u})\delta_{ij} + \eta\left({\partial u_i\over \partial x_j} + {\partial u_j\over \partial x_i}\right)
\label{eq:stressnewtonian}
\end{equation}
in which, $\beta$ is the second viscosity coefficient and $\eta$ is the shear viscosity.

For an incompressible Newtonian fluid, $\nabla\cdot\vec{u}= 0$, the Cauchy momentum equation becomes the Navier-Stokes equations.

\begin{equation}
\rho_F {\partial \vec{u}\over \partial t} + (\rho_F\vec{u}\cdot\nabla) \vec{u}= -\nabla p +  \eta (\nabla\cdot\nabla) \vec{u}  + \rho_F \vec{g}
\label{eq:navierstokes}
\end{equation}

For hydrostatic pressure, $\sigma_{ij}= -p\delta_{ij}$.  For an ideal fluid with constant density and no body forces, and replacing the Eulerian operator, Equation~\ref{eq:cauchy} reduces to the Euler equation.

\begin{equation}
\rho_F{D\vec{u}\over Dt}= -\nabla p
\end{equation}

\subsection{Conservation of energy}

Similarly, with the expressions in Equations~\ref{eq:newpressure}, \ref{eq:newkinetic}, and  ~\ref{eq:newheat}, one can re-write the conservation of energy equation~\cite{RLL1990}.  

\begin{eqnarray}
  &&{\partial  {U/ V}\over \partial t} + \nabla\cdot ( \vec{u} {U/ V} ) + \nabla\cdot \Phi_Q +\nonumber\\
  &&{\partial {1\over 2}\rho_F \left(\vec{u}\cdot\vec{u}\right)\over \partial t} +\nabla\cdot \left( {1\over 2}\rho_F \left(\vec{u}\cdot\vec{u}\right)\right)\vec{u} +\nonumber\\
  &&\nabla\cdot(\overline{\overline{p}}\cdot\vec{u})= \vec{f}_{ext}\cdot \vec{u}
\end{eqnarray}

Replacing Equation~\ref{eq:conservmomentum}.  Using the identity $\nabla\cdot (\overline{\overline{p}}\cdot \vec{u})= (\nabla\cdot \overline{\overline{p}})\cdot \vec{u} + \overline{\overline{p}}:(\nabla\otimes\vec{u})^T=  (\nabla\cdot \overline{\overline{p}})\cdot \vec{u} + \overline{\overline{p}}: \overline{\overline{\nabla\vec{u}}}$ and other vector identities.

\begin{eqnarray}
  &&{\partial  {U/ V}\over \partial t} + \nabla\cdot ( \vec{u} {U/ V} ) + \nabla\cdot \Phi_Q  + \overline{\overline{p}}: \overline{\overline{\nabla\vec{u}}} +\nonumber\\
  &&{\partial {1\over 2}\rho_F \left(\vec{u}\cdot\vec{u}\right)\over \partial t} +\nabla\cdot \left( \left(\vec{u}\cdot\vec{u}\right) {1\over 2}\rho_F\vec{u}\right) =\nonumber\\
  && \left({\partial (\rho_F\,\vec{u})\over \partial t} + 
   \vec{u}(\nabla\cdot \rho_F\vec{u}) + (\rho_F\vec{u}\cdot \nabla)\vec{u}\right)\cdot \vec{u}
\end{eqnarray}
in which,  $\overline{\overline{p}}: \overline{\overline{\nabla\vec{u}}}= \sum_{ij} p_{ij}{\partial u_i/\partial x_j}$ is a scalar.

\begin{eqnarray}
  &&{\partial  {U/ V}\over \partial t} + \nabla\cdot ( \vec{u} {U/ V} ) + \nabla\cdot \Phi_Q  + \overline{\overline{p}}: \overline{\overline{\nabla\vec{u}}}=\nonumber\\
  &&{1\over 2}\vec{u}\cdot \vec{u} \left({\partial \rho_F\over \partial t} + 
   \nabla\cdot (\rho_F\vec{u})\right)
\end{eqnarray}

Replacing Equation~\ref{eq:conservmass}.

\begin{eqnarray}
  &&{\partial  {U/ V}\over \partial t} + \nabla\cdot ( \vec{u} {U/ V} ) + \nabla\cdot \Phi_Q  + \overline{\overline{p}}: \overline{\overline{\nabla\vec{u}}}=\nonumber\\
  &&{1\over 2}(\vec{u}\cdot \vec{u}) {\cal G}_N
\label{eq:conservenergy02}
\end{eqnarray}

\begin{figure} [t]  
\centering
 \includegraphics[width=0.3\linewidth]{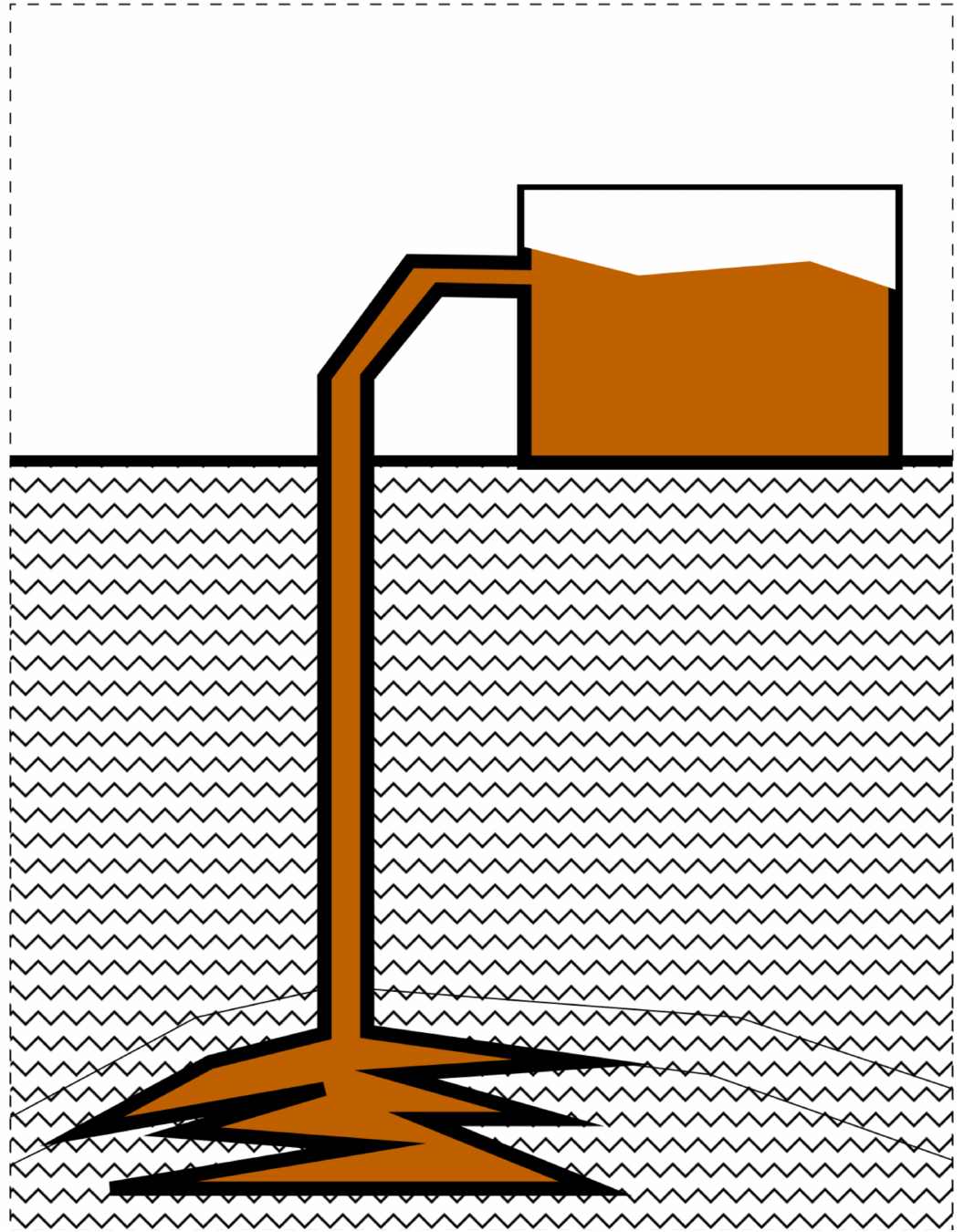}  
\caption{The oil well as a thermodynamic closed system.}
\label{fig4:closedsystem}
\end{figure}

The oil well can be regarded as a thermodynamic system.  The well plus surroundings can be considered a closed system, but the well alone is an open system, which exchanges thermal energy, mass and work with its surroundings, as shown in Figure~\ref{fig4:closedsystem}.  Unfortunately,  the internal energy, $U$,  is not a practical quantity.  Good practical candidates are the enthalpy, $H$, and the Gibbs free energy, $G$ as they are functions of temperature variation, pressure variation, and mass exchange.  Thus, one needs to re-write Equation~\ref{eq:conservenergy02}.  

\begin{eqnarray}
dH&=& TdS + Vdp + \sum_j\mu_jdN_j\\
dG&=& -SdT + Vdp + \sum_j\mu_jdN_j
\end{eqnarray}

Experimentally, it is easier to determine the heat capacity at constant pressure, $C_p$.  Thus, it is more convenient to use the enthalpy.  Writing the entropy, $S$, as a function of temperature and pressure and applying Maxwell identities.

\begin{equation}
dH= \left.{\partial H\over \partial T}\right|_{p,N} \!\!\!dT + \left.{\partial H\over \partial p}\right|_{T,N} \!\!\!dp + \sum_j\left.{\partial H\over \partial N_j}\right|_{p,T} \!\!\!dN_j
\end{equation}

\begin{eqnarray}
\left.{\partial H\over \partial T}\right|_{p,N}&=& C_p\\
\left.{\partial H\over \partial p}\right|_{T,N}&=& -\mu_{JT}C_p
\end{eqnarray}
in which, $\mu_{JT}= \left.{\partial T\over \partial p}\right|_H$ is the Joule-Thomson coefficient.

It can be shown that the Joule-Thomson coefficient is related to the coefficient of thermal expansion.

\begin{equation}
\mu_{JT}= {1\over C_p}\left(T\left.{\partial V\over \partial T}\right|_{p,N} - V\right)= {\alpha_pT -1\over \rho_F c_p}
\end{equation}
in which, $\alpha_p= {1\over V}\left.{\partial V\over \partial T}\right|_{p,N}$ is the coefficient of thermal expansion at constant pressure, and $c_p= C_p/m_F$ is the specific heat capacity.

One can define the enthalpy density to evaluate the exchange of thermal energy.

\begin{equation}
  H_V= H/V= U/V + p + \sum_j\mu_jn_j
\end{equation}
in which, $n_j= N_j/V$.

Dividing the enthalpy density by the fluid density, one gets the specific enthalpy, which is the enthalpy per unit mass.

This is also known as the Local Instant Formulation combined with the Boltzmann Statistical Average~\cite{IH2006}.  The three conservation equations are coupled and must be solved simultaneously.  Unfortunately, this is not an easy task.  Thus, empirical models and assumptions are introduced to reduce the level of coupling.  



\section{Quasi 3D case - transverse averaging technique}
\label{sec:quasi3d}

Defining the average of a scalar, $\phi(\vec{r},z)$, and the average of a vector, $\vec{V}(\vec{r},z)$, over the cross-section area for a axially symmetric wellbore~\cite{BS2007,BS2011}.  Differently from the local instant formulation~\cite{IH2006}, there is no need to assume the area is constant along the longitudinal direction.

\begin{equation}
\overline{\bf \phi}(z)= {1\over A(z)}\int_{A(z)}\phi(\vec{r},z) dS
\label{eq:TATscalar}
\end{equation}

\begin{equation}
\overline{\bf V}(z)= {1\over A(z)}\int_{A(z)}\vec{V}(\vec{r},z) dS
\label{eq:TATvector}
\end{equation}

\begin{figure} [t]  
\centering
 \includegraphics[width=0.3\linewidth]{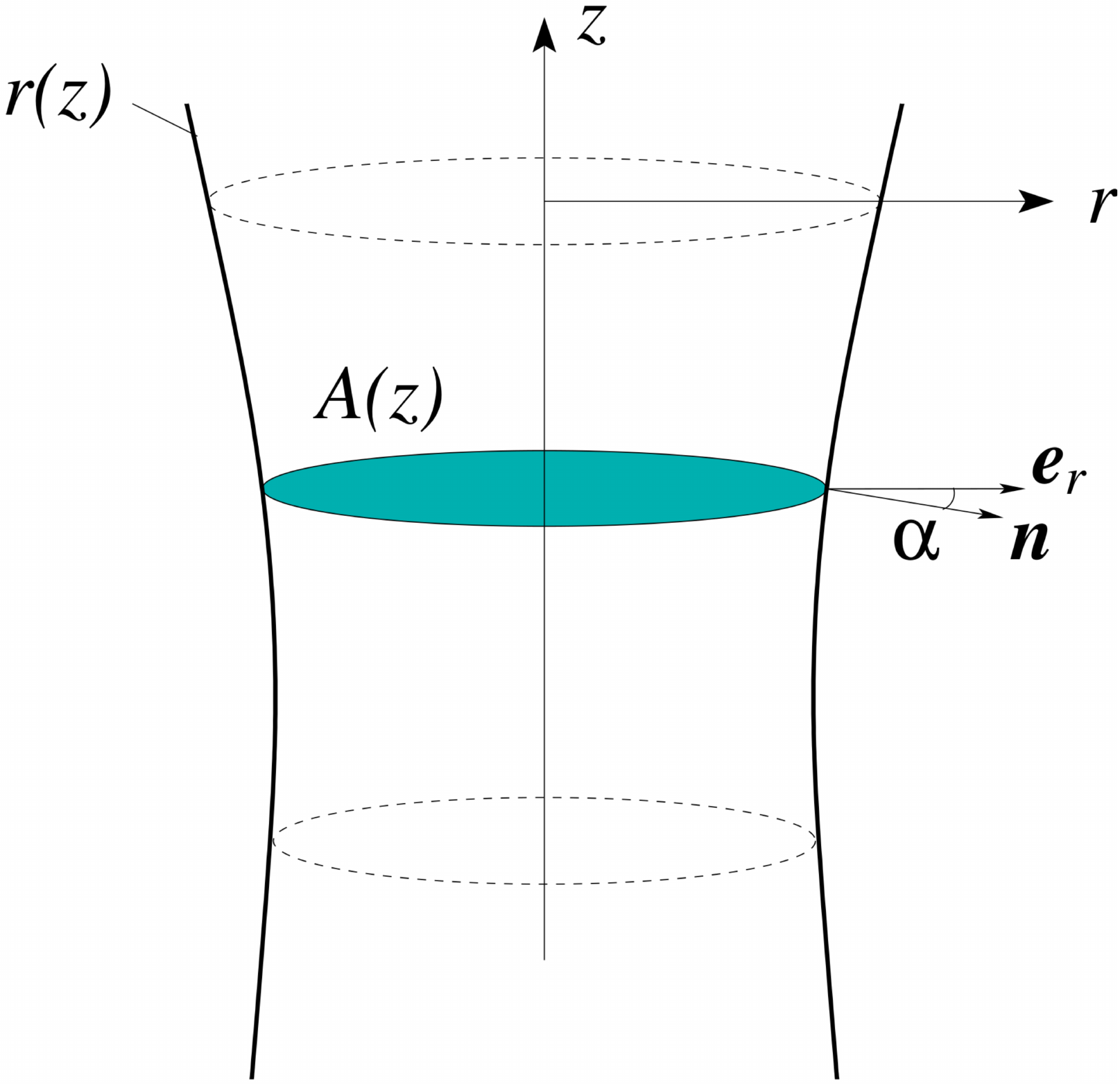}  
\caption{Wellbore profile to apply the transverse averaging technique.}
\label{fig:TAT}
\end{figure}

It is more convenient to use the normal, $\hat{n}(z)$, to the profile curve, $r(z)$, as shown in Figure~\ref{fig:TAT}.

\begin{equation}
\tan\alpha= {dr(z)\over dz} \Longrightarrow \cos\alpha= {1\over \sqrt{1+(dr(z)/dz)^2}}
\end{equation}

Thus, one gets the following identities for the gradient and divergence average~\cite{BS2007,BS2011}.

\begin{equation}
\int_{A(z)} \!\!\!\nabla\phi\,\, dS= {\partial \over \partial z}\left[\hat{e}_z\overline{\bf \phi}(z)A(z)\right] + \oint_{L(z)} {\phi   \hat{n}\over\cos\alpha}dl
\label{eq:TATgrad}
\end{equation}

\begin{equation}
\int_{A(z)}\!\!\!\!\!\nabla\cdot\vec{V}\,\,dS\!\!=  {\partial \over \partial z}\left[\hat{e}_z\!\cdot\!\overline{\bf V}(z)A(z)\right]\!  +\! \oint_{L(z)}\!\!  {\vec{V} \cdot \hat{n}\over \cos\alpha}dl
\label{eq:TATdiv}
\end{equation}

The transverse averaging technique can be applied to the divergence of rank-2 tensors.  To achieve this, one can start with the divergence of a vector, shown in Equation~\ref{eq:TATdiv}, to obtain the following expression.

\begin{equation}
\int_{A(z)}\!\!\!\!\!\!\!\left.\nabla\cdot\overline{\overline{\sigma}}\right|_idS\!\!=  {\partial \over \partial z}\left[\hat{e}_z\!\cdot\!\left.\overline{\sigma}\right|_{i}A(z)\right]\!  +\! \oint_{L(z)}\!\!  {\left.\overline{\overline{\sigma}}\right|_i \cdot \hat{n}\over \cos\alpha}dl
\label{eq:TATdivtensor}
\end{equation}

Using such definitions, the conservation of particles equation, the conservation of momentum equation, and the conservation of energy equation can be re-written to calculate flow rate, pressure profile and temperature profile along an oil well.  Next, the equations are obtained for a single phase in the oil flow.

\subsection{Conservation of particles}

For laminar flow,  one can assume that $(\rho_F\vec{u})\cdot\hat{n}= 0$, i.e., there is no flow in the perpendicular direction to the tubing.  Integrating the equation of conservation of particles over the cross-section area, one can apply the transverse averaging technique, as defined in Equation~\ref{eq:TATscalar} and \ref{eq:TATvector}, and use the identity in Equation~\ref{eq:TATdiv}.

\begin{eqnarray}
  &&{\partial (\overline{\rho}_FA(z))\over \partial t}+{\partial ( \overline{\rho_F{\bf u}}_zA(z))\over \partial z}+\nonumber\\
  &&\cancelto{0}{\oint_{L(z)}\!\!  {( \rho_F \vec{u})\cdot \hat{n}\over \cos\alpha}dl}
  = \overline{\cal G }_N  A(z)
\end{eqnarray}

The source, $\overline{{\cal G }}$, can be used to model the oil flow from the reservoir at the well bottom, and the sink, $\overline{{\cal R }}$, can be used to model the well top, as the oil is stored in a tank.  Should be desired, this equation can be written in terms of the volumetric flow rate, ${\cal Q}_V= \overline {{\bf u}}_zA$ or mass flow rate, ${\cal Q}_M= \overline{\rho_F{\bf u}}_zA$.

Expanding the average of the product in terms of the variation around the average, and noting that $\overline{\delta {\bf u}}_z= \overline{\delta \rho}_F= 0$.  

\begin{equation}
\overline{\rho_F{\bf u}}_z= \overline{\rho}_F\overline{\bf u}_z + \overline{\delta \rho_F\delta {\bf u}}_z 
\end{equation}

The simplest approximation is to neglect second order effects, thus $\overline{\rho_F{\bf u}}_z\approx \overline{\rho}_F\overline{\bf u}_z$.

\begin{equation}
{\partial (\overline{\rho}_FA(z))\over \partial t} +{\partial ( \overline{\rho}_F\, \overline{{\bf u}}_zA(z))\over \partial z}= \overline{\cal G }_N A(z)
\label{eq:massbalance01}
\end{equation}

Assuming the cross-section area is constant along the $z$-direction.

\begin{equation}
{\partial \overline{\rho}_F\over \partial t}+\overline{{\bf u}}_z{\partial  \overline{\rho}_F\over \partial z}+\overline{\rho}_F {\partial  \overline{\bf u}_z\over \partial z}= \overline{\cal G }_N 
\end{equation}

\begin{equation}
{D \overline{\rho}_F\over D t} +\overline{\rho}_F {\partial  \overline{\bf u}_z\over \partial z}= \overline{{\cal G }}_N
\end{equation}

 Defining compressibility at constant temperature, $\beta_T$.

\begin{equation}
\beta_T=   {1\over\overline{\rho}_F}{D\overline{\rho}_F\over D p}=-{1\over V}\left.{\partial V\over \partial p}\right|_T
\end{equation}

The density variation can be replaced with the pressure variation. 

\begin{equation}
\beta_T{D p\over D t} + {\partial \overline{\bf u}_z  \over \partial z}=  {1\over\overline{\rho}_F}\overline{{\cal G }}_N
\label{eq:massbalance02}
\end{equation}

Using the bulk modulus definition,  $K_b=  1/\beta_T$, finally, one gets.

\begin{equation}
{D p\over D t} +  c_a^2\overline{\rho}_F{\partial  \overline{\bf u}_z \over\partial z}= c_a^2\overline{{\cal G }}_N
\label{eq:balancemassdyn}
\end{equation}
in which, $c_a^2= {K_b/ \overline{\rho}_F }$ is the square of the speed of sound.

\subsection{Conservation of momentum}

Calculating the transverse averaging of Equation~\ref{eq:momconserv04}.

\begin{eqnarray}
{\partial  \overline{\rho_F{\bf u}}_zA(z) \over \partial t} &+&  \left.\overline{\nabla\cdot(\rho_F \vec{u}\vec{u}^T )}\right|_zA(z)=\nonumber\\
&&   \left.\overline{\nabla\cdot \overline{\overline{\sigma}}}\right|_zA(z)\, +\, \overline{\bf f}_{ext,z}A(z)
\end{eqnarray}

Assuming there is no momentum flow in the radial direction, and that the radial stress component is assumed constant along the perimeter.

\begin{equation}
\oint_{L(z)}\!\!  {\left.\overline{\overline{\sigma}}\right|_z \cdot \hat{n}\over \cos\alpha}dl= {\tau_{rz}\over \cos\alpha}2\pi r(z)
\end{equation}

Thus. 

\begin{eqnarray}
{\partial  \overline{\rho_F {\bf u}}_zA(z) \over \partial t}  &+&  {\partial \over \partial z}\left[\hat{e}_z\!\cdot\!\left.\overline{\rho_F {\bf uu}}\right|_{z}A(z)\right]+\nonumber\\
&&\cancelto{0}{\oint_{L(z)}\!\!  {\left.\rho_F \vec{u}\vec{u}^T\right|_z \cdot \hat{n}\over \cos\alpha}dl}= \nonumber\\
&&{\partial \over \partial z}\left[\hat{e}_z \cdot\left.\overline{\overline{\sigma}}\right|_{z}A(z)\right] +\nonumber\\
&&{\tau_{rz} 2\pi r(z)\over \cos\alpha} + \overline{\bf f}_{ext,z}A(z)
\end{eqnarray}

Also, neglecting higher order effects, this equation can be further simplified.

\begin{eqnarray}
  {\partial  \overline{\rho}_F \overline{\bf u}_zA(z) \over \partial t} &+&  {\partial \over \partial z}\left[\overline{\rho}_F \overline{\bf u}_z\overline{\bf u}_z \,A(z)\right]= {\partial \over \partial z}\left[\overline{\sigma}_{zz}A(z)\right] +\nonumber\\
  &&   {\tau_{rz} 2\pi r(z)\over \cos\alpha} + \overline{\bf f}_{ext,z}A(z)
  \label{eq:consmomentumTAT}
\end{eqnarray}

\subsection{Conservation of energy}

Finally, the transverse averaging technique is applied to the conservation of energy as described by Equation~\ref{eq:conservenergy02}.  There is exchange of energy, but no exchange of mass along the tubing, this results in the following equation.

\begin{eqnarray}
{\partial \overline{\bf U}A(z)/V\over \partial t} &+& {\partial \over \partial z}\left[\overline{{\bf u}}_z \overline{{\bf U}}A(z)/V\right]+
\overline{\bf p}_{zz}A(z){\partial \overline{\bf u}_z\over \partial z}+
\nonumber\\
&& {\partial \over \partial z}\left[\hat{e}_z\cdot\overline{{\bf \Phi}}_QA(z)\right]+\oint_{L(z)}\!\!  {\Phi_Q \cdot \hat{n}\over \cos\alpha}dl=\nonumber\\
&& {1\over 2}\overline{\bf u}_z^2\overline{\cal G}_NA(z)
\end{eqnarray}

The flow of internal energy in the radial direction is zero.

\begin{equation}
\oint_{L(z)}\!\!  {\left. \overline{\bf U}A(z)/V  \right|_z \cdot \hat{n}\over \cos\alpha}dl= 0
\end{equation}

To extract the temperature profile, it is more convenient to replace the internal energy with the enthalpy density, $\overline{\bf H}_V= \overline{\bf H}/V$.  The enthalpy variation is a measure of thermal energy flow.

\begin{equation}
\overline{\bf H}_V= \overline{\bf U}/V + p + \sum_i\mu_i \overline{\bf n}_i
\end{equation}
in which, $\overline{\bf n}_i= \overline{\bf N}_i/V$.

Resulting in the following equation.

\begin{eqnarray}
  {\partial \overline{\bf H}_VA(z)\over \partial t} &+& {\partial (\overline{{\bf u}}_z \overline{\bf H}_VA(z))\over \partial z}- {\partial pA(z)\over \partial t}-\overline{\bf u}_z {\partial p A(z)\over \partial z}-\nonumber\\
  &&  \sum_i{\partial (\mu_in_iA(z))\over \partial t} -  \sum_i{\partial \over \partial z}\left[\mu_i\overline{{\bf u}}_z n_iA(z)\right]+\nonumber\\
&&{\partial (\hat{e}_z\cdot\overline{{\bf \Phi}}_QA(z))\over \partial z}+\oint_{L(z)}\!\!  {\Phi_Q \cdot \hat{n}\over \cos\alpha}dl=\nonumber\\
  &&  {1\over 2}\overline{\bf u}_z^2\overline{\cal G}_NA(z)
  \label{eq:conservenergyTAT}
\end{eqnarray}

Along the tubing, only thermal energy is exchanged, and it is assumed constant along the perimeter.  Notice that the thermal energy transfer is positive flowing into the tubing.

\begin{equation}
\oint_{L(z)}\!\!  { \Phi_Q \cdot \hat{n}\over \cos\alpha}dl= -{\Phi_{Q,r}\over \cos\alpha}2\pi r(z)
\end{equation}



\section{Discussion}

One can apply the concepts introduced in the previous section to evaluate the one dimensional flow of oil.  This approximation can give an insight in model construction to extract the flow rate, pressure profile and temperature profile. Consider an isotropic homogeneous fluid with constant density in a production well. This is the simplified single phase flow. 

\begin{figure} [t]  
\centering
\includegraphics[width=0.3\linewidth]{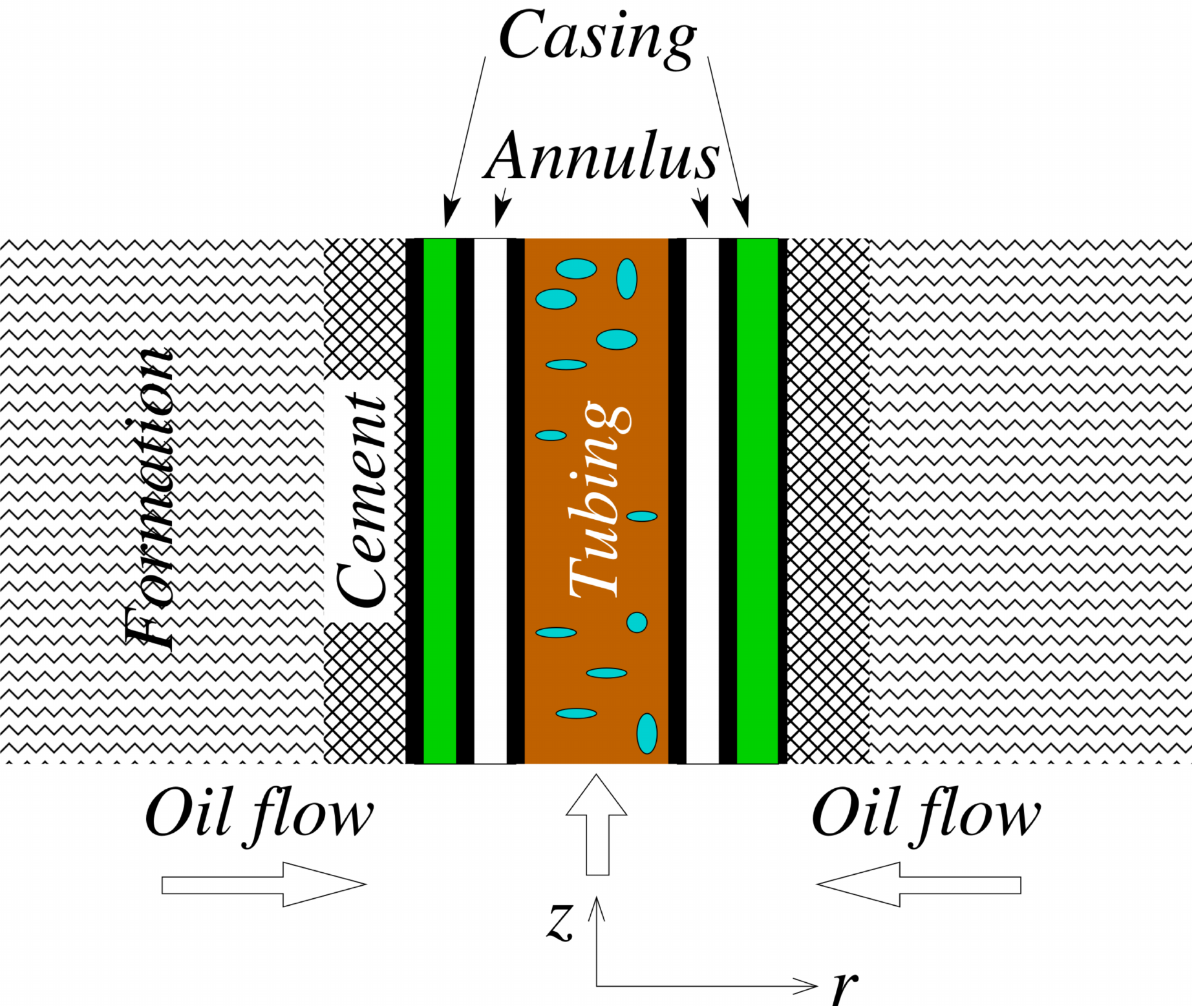}  
\caption{Wellbore structure for the one dimensional approximation.}
\label{fig5:wellbore}
\end{figure}

It is assumed that the oil well tubing displays a constant radius, $r(z)= R$.  Thus, $\cos\alpha= 1$, and $A(z)= \pi R^2$.  As can be seen in Figure~\ref{fig5:wellbore}, oil flows radially from the reservoir inside the drainage radius, $r_d$, into the well.    The external force per unit volume is partly due to the reservoir pressure, which forces the oil upwards, and suffers opposition from gravity.  

The velocity profile, $u_z(r)$, in the tubing is not constant along the radius.  Thus, applying the transverse averaging technique,  one defines the average velocity over the cross-section area.

\begin{equation}
\overline{\bf u}_z= {1\over \pi R^2}\int_0^{2\pi}\int_0^{R} u_z(r) rd\theta dr= \overline{\bf u}
\end{equation}
in which, $u_z(r)$ is the velocity profile along the radius.

One approach is to combine the mass balance and momentum balance equations and solve the energy balance equation separately with the help of thermodynamics to extract the temperature profile~\cite{Izgec2008}.  First, the problem is solved in steady-state.  The state-state solution can be used as initial condition for the dynamic problem.

The estimated temperature profile is used to update temperature dependent properties, which are introduced in the mass conservation and momentum conservation equations to calculate the pressure profile, production flow rate, or sound speed.  Thus, combining the equations obtained in the previous section for the particular case of constant tubing radius, $R$, and constant chemical potential, $\mu_i$, one gets the following set.

\begin{equation}
\left\{
\begin{array}{cl}
  {\partial \overline{\rho}_F\over \partial t} + {\partial (\overline{\rho}_F \overline{\bf u})\over \partial z} &=  \overline{{\cal G}}_N\\
  &\\
  {\partial  \overline{\rho}_F \overline{\bf u} \over \partial t} +  {\partial (\overline{\rho}_F \overline{\bf u}\,\overline{\bf u})\over \partial z}&= {\partial \sigma_{zz}\over \partial z} +  {2 \tau_{rz} \over R} + \overline{\rho}_F \hat{e}_z\cdot\vec{g}\sin{\theta}\\
  &\\
{\partial \overline{\bf H}_V\over \partial t}+{\partial (\overline{\bf u}\overline{\bf H}_V)\over \partial z}&=  {1\over 2}\overline{\bf u}^2 {\cal G}_N + {\partial p\over \partial t}+\overline{\bf u}{\partial p\over \partial z} - \\
&\,\,\,\, {\partial (\hat{e}_z\cdot\overline{{\bf \Phi}}_Q)\over \partial z} + {2\Phi_{Q,r} \over R} +\\
&\,\,\,\,\sum_i\mu_i\left({\partial n_i\over \partial t}+{\partial (\overline{{\bf u}}_z n_i)\over \partial z}\right)\\
\end{array}
\right.
\label{eq:1Dtimedependent}
\end{equation}
in which, $\theta$ is the tubing axis angle with respect to the horizontal and $\overline{{\cal G}}_N$ can be used to model the flow of oil from the reservoir into the well.

 According to Equation~\ref{eq:stressnewtonian}, the stress tensor components for a Newtonian fluid are as follows.

\begin{eqnarray}
 \sigma_{zz}&=&  - p +  {4\over 3} \eta {\partial \overline{\bf u}\over \partial z}=  -p\\
\tau_{zr}&=&  \eta  {\partial \overline{\bf u}\over \partial r}= 0 
\end{eqnarray}
in which, $u_r= 0$.

In general, the multiphase flow in an oil well is not Newtonian, but to a first approximation the single phase flow of light to medium crude oil can be considered Newtonian.  Thus, the simplest approximation for oil/gas mixture flow is to model gas fraction as a compressible flow and oil fraction as a incompressible Newtonian fluid.  To be more accurate, more sophisticated rheological models can be applied, such as: Ostwald-de Waele or power law fluid, Bingham fluid or plastic,  Herschel-Bulkley or viscous-plastic or yield power law, among others.

To account for pressure loss in the pipe due to friction, one can use the Darcy-Weisbach equation.

\begin{equation}
{2 \tau_{zr}\over R}= -f_D{\overline{\rho}_F \overline{\bf u}^2\over 4R} 
\end{equation}
in which,  $f_D= 4f_F= 64/\mbox{Re}$ for $Re<2000$ is the Henry Darcy friction factor, and $f_F$ is the John T. Fanning friction factor.

The third equation is used to calculate the temperature profile.  The laminar flow assumption makes null any convective flow in the fluid layer.   One should keep in mind the quantity for thermal energy flux, as presented in Section~\ref{sec:conservequations} is per unit area.  One can assume that the thermal energy flows mainly by convection in the z-direction as the fluid moves upwards.

\begin{equation}
{\partial (\hat{e}_z\cdot\overline{{\bf \Phi}}_Q)\over \partial z}= -h {\pi R^2\over V} {d T\over d z}
\end{equation}
in which, $h$ is the thermal convection coefficient ($J/(m^2.s.K)$).

In the radial direction, the flow of thermal energy is a combination of conduction and convection.  The first layer is the fluid inside the tubing.  Next, one finds the tubing wall. Then, the annulus, the cement and the sand/rock layers, as shown in Figure~\ref{fig5:wellbore}.  Thus, the radial direction can be modeled as a set of thermal layers, which can be modeled as a cascade of thermal RC-circuit.  In steady-state, only the thermal resistance is considered~\cite{SL2018}.

\begin{equation}
\Phi_{Q}A= \dot{Q}_r= U_\theta A\Delta T= R_{\theta}^{-1}\Delta T
\end{equation}
in which, $\dot{Q}_r$ is the rate of thermal energy transfer, $U_\theta$ is the overall coefficient of thermal energy transfer, and $R_\theta= {1\over U A}$ is the thermal resistance.

\begin{equation}
{1\over U_\theta}= \sum_i{1\over U_{\theta,i}} 
\end{equation}
in which, $ U_{\theta,i}$ is the coefficient of thermal energy transfer for the $i$-th well layer.

\begin{itemize}
\item For linear conduction.

\begin{equation}
R_{\theta}= {\Delta z\over k}{1\over A}
\end{equation}

\item For radial conduction.

\begin{equation}
R_{\theta}= {\ln r_{out}/r_{in}\over 2\pi k \Delta z }
\end{equation}

\item For convection.

\begin{equation}
R_{\theta}= {1\over h} {1\over A}
\end{equation}
\end{itemize}

For temperature transient, one needs to include the thermal capacity of each layer to model the energy transfer as a propagation through a cascade of thermal $R_\theta C_\theta$ circuits.  The obtained set of equations can be used in many flow conditions: no flow, steady-state flow, dynamic flow.



\section{Application}

\subsection{Steady-state: Ramey's model}

The most widely known model for wellbore thermal energy transmission for injection fluids is the Ramey's model~\cite{Ramey1962}.   This model assumes single phase steady-state vertical flow of a incompressible Newtonian fluid, materials properties are invariant with temperature, thermal energy transfer is radial only ($\hat{e}_z\cdot\overline{{\bf \Phi}}_Q= 0$), and the energy flow is much faster than the fluid flow.   The origin, $z= 0$, is set at the wellhead, with the $z$ axis pointing downward.  Under such assumptions Equation~\ref{eq:1Dtimedependent} is simplified as follows.

\begin{equation}
\left\{
\begin{array}{cl}
{\partial \overline{\bf u}\over \partial z} &= 0\\
  &\\
 0&= -{d p\over d z} + \overline{\rho}_F g\\
 &\\
 {d \overline{\bf H}_V\over d z}&= {d p\over d z} + {1\over\overline{\bf u}}\left[{2\Phi_{Q,r} \over R}\right] \\
\end{array}
\right.
\label{eq:rameyset}
\end{equation}

The enthalpy per unit volume can be written as a function of temperature and pressure.

\begin{eqnarray}
  d \overline{\bf H}_V&=& (C_p/V)  d T + (1-\alpha_p T)d p\\
  &=& (C_p/V)  d T - (C_p/V)\mu_{JT}d p
  \label{eq:enthalpydiff}
\end{eqnarray}

For incompressible flow, $\alpha_p= 0$.  Thus, $\mu_{JT}= -{1\over\overline{\bf \rho}_F c_p}= -{V\over C_P}$.

\begin{equation}
  {d \overline{\bf H}_V \over d z}= (C_p/V)  {d T\over d z} + {d p\over d z}
\end{equation}

Thus.

\begin{equation}
  {d \overline{\bf H}_V \over d z}= (C_p/V)  {d T\over d z}=  {1\over\overline{\bf u}}\left[ {2\Phi_{Q,r} \over R}\right] 
\end{equation}

Along the well, the equilibrium temperature at a distant position from the well is set by the geothermal profile.  This is the surroundings temperature, $T_s$.  In the simplest case, the right side is constant, yielding a linear geothermal profile.

\begin{equation}
{d  T_s \over d z}= \mbox{constant}
\end{equation}

\begin{equation}
T_s(z)= az+b
\end{equation}
in which, $a$ is the geothermal gradient, and $b$ is the surface temperature.

As the tubing temperature, $T_t$, is not known, it is best to write this temperature in terms of the surroundings temperature, $T_s$.  To achieve this, one can subdivide the radial flow of thermal energy into two sections in series: tubing to casing and casing to surroundings.

\begin{equation}
(R_{\theta,tc} + R_{\theta,cs})\dot{Q}_r= (T_t - T_c) + (T_c - T_s)
\end{equation}
in which, $\dot{Q}= \Phi_Q\times A$.

\begin{equation}
A_R= \overline{\bf u} C_p \left({R_{\theta,tc} R A\over 2V} + {R_{\theta,cs} R A\over 2V}\right)
\end{equation}

\begin{equation}
A_R=  {\overline{m}_F\overline{\bf u}\over \Delta z} {C_p\over \overline{m}_F}\left({1\over 2 \pi R U_\theta} +  {R_{\theta,cs} A\over 2\pi R}\right)
\end{equation}
in which, $U_\theta$ is the overall coefficient of thermal energy transfer for all well layers from the liquid inside the tubing to the casing.

Defining the fluid density, $\overline{\rho}_F= \overline{m}_F/V$, the fluid injection rate, $w= (\overline{m}_F \overline{\bf u}/\Delta z)=  \overline{\rho}_F \overline{\bf u}A$, and the specific heat at constant pressure, $c_p=  (C_p/\overline{m}_F)$.

\begin{equation}
A_R= w c_p\left({1\over 2\pi R U_\theta}  +  {R_{\theta,cs} A\over 2\pi R}\right)
\end{equation}

According to Ramey's model, a time-dependent dimensionless function $f(t)$ is introduced to account for the thermal energy flow from the casing into the surrounding formation.

\begin{equation}
A_R(t)= w c_p\left({1\over 2\pi R U_\theta} + { f(t)\over 2\pi k}\right)
\end{equation}
in which, $t$ is the injection time period.

The thermal energy transfer is negative as it goes from the tubing to the surroundings. 

\begin{equation}
  {d  T_t \over d z}= {1\over\overline{\bf u}(C_p/V)}\left( {2\Phi_{Q,r} \over R}\right)= - {1\over w c_P}\left( {2\dot{Q}_r\over R}\right)
\end{equation}

\begin{equation}
  {d T_t(z,t)\over d z}= -{T_t(z,t) - T_s(z)\over A_R(t)}= -{T_t\over A_R(t)} + {(az+b)\over A_R(t)}
\end{equation}

Solving this equation with the integrating factor method, one gets the Ramey's model for the tubing temperature profile as a function of time for incompressible Newtonian liquid injection or a modified version for gas injection.

\begin{itemize}
\item Newtonian liquid injection.
  \begin{equation}
    T_t(z,t)= az + b -A_R(t)a + (T_0 - b + A_R(t)a)e^{-z/A_R(t)}
  \end{equation}
  
\item Gas injection.
  \begin{eqnarray}
    T_t(z,t)= az + b - A_R(t)\left(a + {1\over c_w c}\right) +&&\nonumber\\
    \!\!\!\!\!\!\!\!\!\! \left[ T_0 - b + A_R(t)\left(a + {1\over c_w c}\right)\right]e^{-z/A_R(t)}&&
  \end{eqnarray}
  
\end{itemize}
in which, $T_0= T(0,t)$ is the constant injected fluid temperature at the surface, $z= 0$ and $c_w= 4.186\, J.g^{-1}\,^oC^{-1}= 778.24 ft.lb\!f$ is the specific heat of water (mechanical equivalent of heat).

\subsection{Transient: simplified Hasan's model}

Firstly, consider a steady-state Newtonian oil flow~\cite{Izgec2008,HK2012}.  As in Ramey's model the transfer of thermal energy is assumed radial only, $\hat{e}_z\cdot\overline{{\bf \Phi}}_Q= 0$.  The fluid density is not constant, thus Equation~\ref{eq:1Dtimedependent} is reduced as follows.

\begin{equation}
\left\{
\begin{array}{cl}
  {\partial (\overline{\rho}_F \overline{\bf u})\over \partial z} &=  0\\
  &\\
  (\overline{\rho}_F \overline{\bf u})  {\partial \overline{\bf u}\over \partial z}&= - {\partial p\over \partial z}  + \overline{\rho}_F g\sin{\theta}\\
  &\\
{\partial (\overline{\bf u}\overline{\bf H}_V)\over \partial z}&= \overline{\bf u}{\partial p\over \partial z} + {2\Phi_{Q,r} \over R} 
\end{array}
\right.
\end{equation}

According to Equation~\ref{eq:enthalpydiff}, the enthalpy per unit volume can be written as a function of temperature and pressure.  As the equation is presented in international units, SI, there is no need to introduce the $g_c$ conversion factor.  The flow of thermal energy towards the surroundings is accounted as negative.

\begin{equation}
  {d T\over d z}=  \left(\mu_{JT} + {1\over \overline{\rho}_Fc_p}\right){\partial p\over \partial z} - {1\over  w c_p }\left[ {2 \dot{Q}_r \over  R}\right]
  \label{eq:hasanstedy}
\end{equation}

To model the radial thermal energy flow, one can use Ramey's model.

\begin{equation}
  {d T\over d z}=  \left(\mu_{JT} + {1\over \overline{\rho}_Fc_p}\right){\partial p\over \partial z} -{T_t(z,t) - T_s(z)\over A_R(t)}
\end{equation}

Next, considering transient flow~\cite{Izgec2008,HK2012} of a Newtonian fluid, Equation~\ref{eq:1Dtimedependent} is reduced as follows.

\begin{equation}
\left\{
\begin{array}{cl}
  {\partial \overline{\rho}_F\over \partial t} + {\partial (\overline{\rho}_F \overline{\bf u})\over \partial z} &= 0\\
  &\\
  {\partial  \overline{\rho}_F \overline{\bf u} \over \partial t} +  {\partial (\overline{\rho}_F \overline{\bf u}\,\overline{\bf u})\over \partial z}&= - {\partial p\over \partial z}  + \overline{\rho}_F g\sin{\theta}\\
  &\\
{\partial \overline{\bf H}_V\over \partial t}+{\partial (\overline{\bf u}\overline{\bf H}_V)\over \partial z}&=  {\partial p\over \partial t}+\overline{\bf u}{\partial p\over \partial z} + {2\Phi_{Q,r} \over R} 
\end{array}
\right.
\end{equation}

Combining the first two equations.

\begin{equation}
\overline{\rho}_F  {\partial   \overline{\bf u} \over \partial t} + \overline{\rho}_F \overline{\bf u} {\partial \overline{\bf u}\over \partial z}= \overline{\rho}_F  {D  \overline{\bf u} \over D t}= - {\partial p\over \partial z}  + \overline{\rho}_F g\sin{\theta}
\end{equation}

Assuming constant mass flow rate, $w= \overline{\rho}_F \overline{\bf u}A$, and replacing the enthalpy per unit volume for the specific enthalpy, $\overline{\bf H}_m= \overline{\bf H}/m_F$.

\begin{equation}
  \overline{\bf H}_V= {m_F\over  V}{\overline{\bf H}\over m_F}=  \overline{\bf \rho}_F\overline{\bf H}_m
\end{equation}

\begin{equation}
 \overline{\bf u}  \overline{\bf H}_V= {w\over  A}{\overline{\bf H}\over m_F}= {w\over  A}\overline{\bf H}_m
\end{equation}

\begin{equation}
 d  \overline{\bf H}_V=  \overline{\bf \rho}_F c_p dT -  \overline{\bf \rho}_F c_p\mu_{JT} dp 
\end{equation}

\begin{eqnarray}
  {\partial \overline{\bf H}_V\over \partial t} + {w\over A}{\partial \overline{\bf H}_m\over \partial z}&=&  \overline{\bf \rho}_F c_p  {\partial T \over \partial t} -  \overline{\bf \rho}_F c_p\mu_{JT}  {\partial p \over \partial t}  + \nonumber\\
  &&{w c_p\over A}\left(  {\partial T \over \partial z} -  \mu_{JT}  {\partial p \over \partial z} \right)\\
&=&  {\partial p\over \partial t}+\overline{\bf u}{\partial p\over \partial z} + {2\Phi_{Q,r} \over R}
\end{eqnarray}

\begin{eqnarray}
  \overline{\bf \rho}_F c_p  {\partial T \over \partial t} +  {w c_p\over A} {\partial T \over \partial z} &=& \mu_{JT} \overline{\bf \rho}_F  c_p\left(  {\partial p \over \partial t} +   {w\over A \overline{\bf \rho}_F }  {\partial p \over \partial z} \right) +\nonumber\\
&&  {\partial p\over \partial t}+\overline{\bf u}{\partial p\over \partial z} + {2\Phi_{Q,r} \over R}
\end{eqnarray}

\begin{eqnarray}
  {\partial T \over \partial t} +  {w \over A \overline{\bf \rho}_F } {\partial T \over \partial z} &=& \mu_{JT} \left(  {\partial p \over \partial t} +   {w\over A \overline{\bf \rho}_F }  {\partial p \over \partial z} \right) +\nonumber\\
  &&{1\over  \overline{\bf \rho}_Fc_P}\left( {\partial p\over \partial t}+\overline{\bf u}{\partial p\over \partial z}\right)+ \nonumber\\
  &&{1\over  \overline{\bf \rho}_Fc_P}\left({2\Phi_{Q,r} \over R}\right)
\end{eqnarray}

\begin{equation}
{D T\over D t}= \left(\mu_{JT}+{1\over  \overline{\bf \rho}_Fc_P}\right){D p\over D t} + {1\over  \overline{\bf \rho}_Fc_P}\left({2\Phi_{Q,r} \over R}\right)
\end{equation}

The final equation reduces to Equation~\ref{eq:hasanstedy} in steady-state.  Similarly, one can use Ramey's model for the radial thermal energy flow.

\begin{equation}
{D T\over D t}= \left(\mu_{JT}+{1\over  \overline{\bf \rho}_Fc_P}\right){D p\over D t} -{T_t(z,t) - T_s(z)\over A_R(t)}
\end{equation}

\subsection{Two-phase flow}

Two-phase flow is of great importance in many areas~\cite{IH2006,DP1999}.  As an example of two-phase flow, one can consider a mixture of gas phase and liquid phase with liquid holdup, $H_L$.  The simplest model is to write the density as a linear combination.

\begin{equation}
\rho_F= \rho_LH_L + \rho_G(1-H_L)
\end{equation}
in which, $H_L$ is the liquid fraction also known as liquid holdup, and the density is a function of temperature.

An equivalent expression can be written for the fluid velocity.

\begin{equation}
\rho_F \vec{u}= \rho_L\vec{u}_LH_L + \rho_G\vec{u}_G(1-H_L)
\end{equation}

Should the gas and liquid velocity be the same, one has the pseudo-fluid approximation as found in the Homogeneous Equilibrium Model (HEM).  Should the velocities be allowed to differ one has the Separated Flow Model (SFM) or Drift Flux Model (DFM), depending on the closure equations.  

To get the temperature profile, one can use the following approximation.

\begin{eqnarray}
  {1\over A} {\partial \overline{\bf H}_VA\over \partial t}&\approx& \overline{\rho}_F c_p {\partial T \over \partial t} -\mu_{JT} \overline{\rho}_F c_p {\partial p \over \partial t}\\
 {1\over A}{\partial ( \overline{\bf u}\overline{\bf H}_VA)\over \partial z} &\approx&\overline{\rho}_F c_p \overline{\bf u}{\partial T \over \partial z} -\mu_{JT}\overline{\rho}_Fc_p\overline{\bf u} {\partial p \over \partial z}
\end{eqnarray}

Above the reservoir there is no flow from the surroundings into the tubing, thus $\overline{{\cal G}}_N= 0$.  From Equations~\ref{eq:massbalance01}, \ref{eq:consmomentumTAT} and \ref{eq:conservenergyTAT}, one can see that new terms are obtained should the area change with time or position.

\begin{equation}
\left\{
\begin{array}{cl}
  {\partial \overline{\rho}_F\over \partial t}&= -{\partial \over \partial z}\left(\overline{\rho}_F\overline{\bf u}\right)- \boxed{\overline{\rho}_F{D\ln A\over Dt}}\\
  &\\
   {\partial p\over \partial z}&= - {D  \overline{\rho}_F \overline{\bf u} \over D t} - \overline{\rho}_F \overline{\bf u} {\partial \overline{\bf u}\over \partial z} +\\
   &{2 \tau_{rz} \over r(z)\cos\alpha} + \overline{\rho}_F \hat{e}_z\cdot\vec{g}\sin{\theta}-\\
   & \boxed{\overline{\rho}_F\overline{\bf u}{D\ln A\over Dt}} - \boxed{p{\partial\ln A\over \partial z}}\\
  &\\
{D T \over D t}&= \left(\mu_{JT}  + {1\over  \overline{\bf \rho}_Fc_P}\right) {D p\over D t}- \\
& {1\over A \overline{\bf \rho}_Fc_P}{\partial (\hat{e}_z\cdot\overline{{\bf \Phi}}_QA)\over \partial z} +  {1\over  \overline{\bf \rho}_Fc_P} {2 \Phi_{Q,r} \over r(z)\cos\alpha }+\\
&\boxed{{p\over  \overline{\bf \rho}_Fc_P}{D \ln A\over D t}} 
\end{array}
\right.
\label{eq:twophase}
\end{equation}

Examples of area models.
\begin{itemize}
\item Constant area, $A= A_0$.
\item Exponential deposition, $A= A_0e^{-t\over \tau}$.
\item Power law, $A= A_0\left(\sum_{n= 0}^N \left({z\over L}\right)^n\right)$.
\item Combined,  $A= A_0\left(\sum_{n= 0}^N \left({z\over L}\right)^n\right)e^{-t\over \tau}$.
\end{itemize}

The model equations are combined with equations of state, empirical equations, and ancillary parameters~\cite{JM2004}.  It can also be applied under the assumptions of the Two Fluid Model (TFM) by writing one set of equations for each phase and including equations for momentum and energy transfer between phases.



\section{Conclusion}

The Transverse Averaging Technique (TAT) is applied to the conservation equations for particles, momentum and energy to provide the 1D+2D approximation in formal grounds.   In this work, TAT is extended to tensor quantities.  Within this approximation technique, a systematic model construction for oil and gas flow is presented, which can be used to examine the various contributions and to calculate analytic expressions for a variety of problems.   The equations obtained for the flowrate and pressure profile are equivalent to the equations obtained by other methods, such as HEM, except that the generation term is maintained.  The net generation term can be used to model the oil flow from the reservoir into the tubing.  The energy equation is presented in terms of enthalpy, which is easier to relate to the temperature.

The TAT technique can be applied to single phase flow or to two-phase flow.  For single phase flow, two classical models are obtained within this approximation, namely: Ramey's steady-state model, and Hasan's transient model.  For the two-phase flow, it can be applied under the assumptions of the  Homogeneous Equilibrium Model (HEM), Separated Flow Model (SFM), Drift Flux Model (DFM), or Two Fluid Model (TFM).   The proposed quasi-3D model is useful to obtain analytic solutions which provide physical insight into the phenomena under scrutiny, including the validation of software tools.  As can be seen in Equation~\ref{eq:twophase}, new terms are obtained should the area of the tubing vary analytically or randomly along the longitudinal direction, as a result of design, deposits, tubing roughness or other. 

\section*{Acknowledgment}

This work is dedicated to Prof. Anatoly A. Barybin of Saint-Petersburg State Electrotechnical University, {\it in memoriam}.

The author thanks CNPq for its continued support.  He also thanks eng. Maur\'{\i}cio Galasssi and eng. Manoel Feliciano da Silva Jr., both from PETROBRAS Brazilian Oil Company for bringing this problem to his attention.

\section*{Symbol nomenclature}

A single bar over the quantity indicates transverse area averaging.  A double bar over the quantity indicates it is a tensor.

\begin{itemize}
\item[] $\alpha$ -- angle between the local normal and the radial direction.
\item[] $\alpha_p$ -- coefficient of thermal expansion.
\item[] $\beta_T$ -- compressibility at constant temperature.
\item[] $\Phi_Q$ -- flux of thermal energy density.  
\item[] $\Phi_q$ -- total flux of kinetic energy density. 
\item[] $\rho_F$ -- fluid density.
\item[] $\rho_G$ -- gas phase density.
\item[] $\rho_L$ -- liquid phase density.
\item[] $\overline{\overline{\sigma}}$ -- Cauchy stress tensor.
\item[] $\sigma_{ii}$ -- normal stresses.
\item[] $\tau_{ij}$ -- shear or viscous stresses.
\item[] $\mu_j$ -- chemical potential for the $j$-phase.
\item[] $\mu_{JT}$ -- Joule-Thomson coefficient.
\item[] $A$ -- tubing cross-section area. 
\item[] $C_p$ -- heat capacity at constant pressure.
\item[] $C_\theta$ -- thermal capacity.
\item[] $c_a$ -- acoustic speed.
\item[] $c_p$ -- specific heat capacity at constant pressure.
\item[] $c_w$ -- specific heat of water. 
\item[] $e_K$ -- total kinetic energy density.
\item[] $F_B$ -- probability density function or distribution function or particle density function.
\item[] $f_D$ -- Henry Darcy friction factor.
\item[] $f_F$ -- John T. Fanning friction factor.
\item[] $\vec{f}_{ext}$ -- external force vector per unit volume.
\item[] $K_b$ -- bulk modulus.
\item[] $k$ -- thermal conduction coefficient.
\item[] $G$ -- Gibbs free energy.
\item[] ${\cal G}_N$ -- net generation. 
\item[] $H_m$ -- enthalpy per unit mass or specific enthalpy.
\item[] $H_V$ -- enthalpy per unit volume.
\item[] $H_L$ -- liquid fraction or holdup.
\item[] $h$ -- thermal convection coefficient.
\item[] $m$ -- parcel mass.
\item[] $\overline{\overline{P}}$ -- total pressure tensor or total flux of momentum density. 
\item[] $\overline{\overline{p}}$ -- pressure tensor.
\item[] $\dot{Q}_r$ -- rate of thermal energy transfer.
\item[] $R_\theta$ -- thermal resistance.
\item[] $r(z)$ -- tubing cross-section radius as a function of the longitudinal direction.
\item[] $T$ -- temperature.
\item[] $T_c$ -- casing temperature.
\item[] $T_s$ -- surroundings temperature.
\item[] $T_t$ -- tubing temperature.
\item[] $U$ -- internal energy.
\item[] $U_\theta$ -- coefficient of thermal resistance.
\item[] $\vec{u}$ -- average parcel velocity or fluid velocity.
\item[] $\vec{u}_G$ --  average gas fraction velocity.
\item[] $\vec{u}_L$ --  average liquid fraction velocity.
\item[] $V$ -- volume.
\item[] $\vec{v}$ -- fluid parcel velocity.
\item[] $w$ -- mass flow rate.
\end{itemize}

\bibliographystyle{model1-num-names}
\bibliography{<your-bib-database>}

\vfil

\end{document}